\documentstyle[twoside,fleqn,espcrc2,epsf]{article}

\newcommand{\AmS}{{\protect\the\textfont2
  A\kern-.1667em\lower.5ex\hbox{M}\kern-.125emS}}

\title{\vspace{-5.0cm} 
\begin{flushright}
{\normalsize Poster presented at ``Lattice 98'' international
symposium, July 13-18, 1998, Boulder, CO, USA}\\
\vspace{-0.2cm}
{\normalsize RIKEN BNL Research Center preprint}\\
\vspace{-0.2cm}
{\normalsize BNL-HET-98/28}\\   
\vspace{-0.2cm}
{\normalsize KEK-TH-587}\\   
\end{flushright}
\vspace*{2.5cm}
SU(4) pure-gauge string tensions}

\author{Shigemi Ohta\address{Institute for Particle and Nuclear
	Studies, KEK, Tsukuba, Ibaraki 305-0801, Japan}
        \thanks{SO thanks the hospitality of the RIKEN BNL Research
	Center where he stayed for a year from September 1997.  This
	research started during this stay.}
        and 
        Matthew Wingate\address{RIKEN BNL Research Center, Brookhaven
	National Laboratory, Upton, NY 11973-5000, USA}}
       
\begin{document}

\begin{abstract}
In response to recently renewed interests in SU(\(N\)) pure-gauge
dynamics with large \(N\), both from M/string duality and from
finite-temperature QCD phase structure, we calculate string tensions
acting between the fundamental \({\bf 4}\), diquark \({\bf 6}\) and
other color charges in SU(4) pure-gauge theory at temperatures below
the deconfining phase change and above the bulk phase transition.  Our
results suggest \({\bf 4}\) and \({\bf 6}\) representations have
different string tensions, with a ratio of \(\sigma_6/\sigma_4\sim
1.3\).  We also found the deconfining phase change is not strong.
\end{abstract}

\maketitle

There are renewed interests in pure-gauge dynamics with SU(\(N\))
gauge groups with \(N>3\): New developments in M/string theory like
Maldacena's duality conjecture \cite{Maldacena} map super conformal
field theory spectra to those in large-\(N\) and strong-coupling
pure-gauge theory \cite{Strassler}.  Another line of thought is
questioning how well the conventional large-\(N\) expansion describes
quenched SU(3) deconfining and related phase transitions or changes
\cite{Rob}.  And there is a long-standing problem of how hadronic
string tension is related to finite-temperature phase structure of
hadrons.

From a M/string-provoked viewpoint, an interesting quantity to look at
is the ratio of string tensions.  In SU(2) and SU(3) pure-gauge
theories there cannot be different string tensions: any irreducible
representation is screened by gluons (in the adjoint representation)
down to either singlet or fundamental representations: in SU(3),
\({\bf 6}\otimes {\bf 8} = {\bf 24} \oplus {\bf 15} \oplus {\bf 6}
\oplus {\bf \bar{3}}\).  This has been confirmed by a numerical
Monte-Carlo calculation \cite{OFU}.  However SU(\(N\)) gauge groups
with \(N>3\) can allow different string tensions for different
representations.  For example in SU(4), the \({\bf 6}\)- and \({\bf
10}\)-dimensional diquark representations cannot be screened by the
\({\bf 15}\) gluon down to the \({\bf 4}\) fundamental: \({\bf
6}\otimes {\bf 15} = {\bf 64} \oplus {\bf 10} \oplus {\bf \bar{10}}
\oplus {\bf 6}\) and \({\bf 10}\otimes {\bf 15} = {\bf 70} \oplus {\bf
64} \oplus {\bf 10} \oplus {\bf 6}\).  So this theory allows at least
two different string tensions, one for \({\bf 4}\) and another for
\({\bf 6}\) representations.  For general \(N\) these string tensions
are classified by the \(N\)-ality \(k\) in the center Z(\(N\)) group
\cite{Strassler}: the fundamental representation has \(N\)-ality
\(k=1\), the diquark representations \(k=2\), and the adjoint \(k=0\),
etc.  One would expect \(\sigma_{k=0} = 0\) and \(1 \le \sigma_{k\ne0}
/ \sigma_{k=1} \le k\).  Indeed naive strong coupling expansion gives
the ratio \(\sigma_k/\sigma_{k=1} = \min\{k, N-k\}\) from multiple
tiling of large Wilson loops.  More sophisticated analysis suggests
\(k(N-k)/(N-1)\) for the non-super-symmetric case or
\(\sin(k\pi/N)/\sin(\pi/N)\) for weakly broken \({\cal N}\)=2
super-symmetry \cite{Strassler}.

From the more conventional large-\(N\) expansion, an important
question yet to be resolved is the weakness of the deconfining phase
transition in the SU(3) pure-gauge thermodynamics \cite{Columbia}.
This phase transition separates the confining phase in which the
center Z(3) symmetry is preserved from the deconfining phase in which
the symmetry is broken.  Conventional large-\(N\) analysis would give
the following picture \cite{Rob}: 1) The transition is located at a
temperature \(T_d \sim O(1)\).  2) it separates the confining phase
with free energy \(F \sim O(1)\) and the deconfining phase with \(F
\sim O(N^2)\).  So the phase transition would be a strongly
first-order one with a latent heat of \(\sim O(N^2)\).  It should not
be affected by finite \(N_f\) flavors of light quarks if \(g^2N\) is
held fixed.  Thus large-\(N\) would not seem to be a good guide for
\(N=3\) thermodynamics where the pure-gauge first-order phase
transition is weak and can be washed away by a few light quark
flavors.  On the other hand if large-\(N\) were to give second-order
phase transition, as suggested by the disappearance of center symmetry
there \cite{Greensite}, then it would still serve as a useful guide:
in this case the weak first-order phase transition of SU(3) pure-gauge
theory is just a peculiarity arising from the cubic effective
interaction present only in this case.

Also interesting is the ratio of the deconfining phase transition
temperature and the fundamental string tension.  In the classic
Hagedorn analysis the ratio should be \(\sqrt{3/\pi(d-2)}\) in \(d\)
space-time dimensions.  As is well known at \(d\)=4 and \(N\)=2 and 3
the actual ratio are smaller than this prediction, and so are in
\(d\)=3 \cite{Karsch}. 

In the present work we study pure-gauge SU(4) dynamics with
single-plaquette action in the fundamental \({\bf 4}\) representation.
We use combinations of Metropolis or pseudo-heatbath update and
over-relaxation update algorithms.  Our lattices are \(\mbox{\rm 4 or
6} \times (\mbox{\rm 4, 6, 8, 12, or 16})^3\) at more than a dozen
values of inverse-squared coupling in \(10.10 \le \beta = 8/g^2 \le
11.00\) with at least 1000 sweeps and up to 8000 near the
deconfinement phase change.  We impose periodic boundary condition in
all the four axes in order to look at finite-temperature systems.

We calculate the plaquette and Polyakov lines in \({\bf 4}\)
(fundamental), \({\bf 6}\) (anti-symmetric diquark), \({\bf 10}\)
(symmetric diquark) and \({\bf 15}\) (adjoint) representations, the
deconfinement fraction derived from them \cite{df} and correlations of
Polyakov lines in all the mentioned representations.  We use the PPR
multihit method \cite{multihit} to reduce noise.

We confirmed the known bulk phase transition \cite{bulk} at around
\(\beta \sim 10.20\) where fluctuation in both plaquette and
deconfinement fraction becomes large.  This is a phase transition that
separates two different confining phases with different correlation
lengths.  The bulk nature of this transition is known because its
position in \(\beta\) does not move as we change the temperature
lattice size from \(L_t\)=4 to 6.  However with \(L_t\)=4 the change
of correlation length drives deconfinement.  So we choose \(L_t\)=6
for the rest of our calculations.

\begin{figure}
\epsfxsize=70mm
\leavevmode
\epsffile[109  225  533  549]{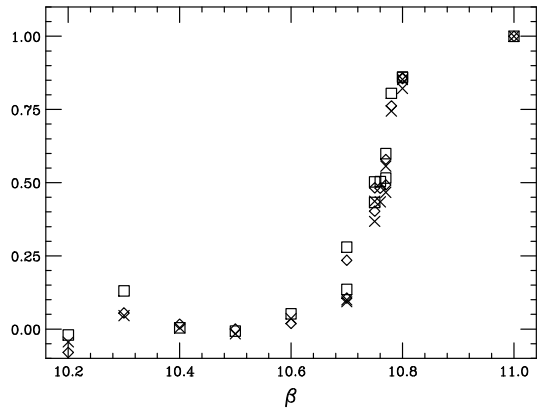}
\caption{Deconfinement fraction from \protect\(6 \times 12^3\)
lattices.  Varying the threshold angle $\theta$ does not change the
determination of the phase change location: \protect\(\theta =
20^\circ\protect\) (cross), \protect\(25^\circ\protect\) (diamond),
\protect\(30^\circ\protect\) (square), hot and cold starts.
\protect\(15^\circ\protect\) and \protect\(35^\circ\protect\) results
look similar and are omitted for simplicity.}
\label{fig:df}
\end{figure}

The average Polyakov line is the order parameter of the center Z(4)
symmetry.  In the confining phase it vanishes, while in the
deconfining phase it acquires a finite value in one of the four Z(4)
(real and imaginary) axes for \({\bf 4}\) and \({\bf 10}\) representations, 
and one of the two Z(2) (real) axes for \({\bf 6}\) and \({\bf 15}\) which
are self-dual representations.  At temperatures lower than
\(\beta\)=10.7, all these Polyakov lines cluster around the origin.
Above that temperature they start to fluctuate along the allowed axes
and gradually deviates from the origin.  At temperatures above
\(\beta=11.0\) they have finite values in one of the allowed axes.
The deconfinement fraction behaves in accord: it starts to deviate
from zero at \(\beta\)=10.7, crosses the 50\% mark at about 10.75 and
reaches 100\% by 11.0 for all threshold angle between \(15^\circ\) and
\(35^\circ\) (see Figure \ref{fig:df}).  From these we conclude that
there is a finite-temperature deconfining phase change at around
\(\beta\)=10.75 for \(L_t\)=6.
On the other hand we do not find any discontinuity in the average
plaquette, and especially in the difference of spacelike and timelike
plaquettes, in this temperature range.  In other words we are not
finding any evidence for a first-order phase transition for \(L_t\)=6
and volumes upto \(16^3\).  This does not exclude first-order
deconfining phase transition on larger volumes, but it would be a weak
one at best \cite{Columbia}.

\begin{figure}
\epsfxsize=70mm
\leavevmode
\epsffile[143  228  456  549]{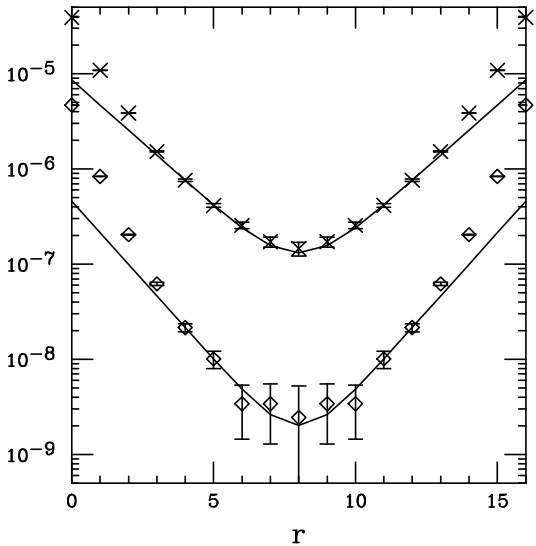}
\caption{Polyakov line correlation in the fundamental
(\protect\({\bf 4}\protect\), cross) and anti-symmetric diquark
(\protect\({\bf 6}\protect\), diamond) representations at
\protect\(\beta=10.65\protect\) on a \protect\(6\times 16^3\protect\)
lattice with 800 configurations.  The curves are fits to tensions of
\protect\(\sigma_4\protect\)=0.102 and
\protect\(\sigma_6\protect\)=0.128.}
\label{fig:tension}
\end{figure}

The correlations of dual Polyakov lines for irreducible
representations \(i\)=\({\bf 4}\), \({\bf 6}\), \({\bf 10}\) and
\({\bf 15}\), etc, are related to the free energy \(F_i(r)\) of a pair
of infinitely heavy dual color charges in the respective
representations separated by the distance \(r=|\vec{r}|\):
\[
C_i(r)=\langle L_i(\vec{x}) L_i^*(\vec{x}+\vec{r}) \rangle_{\vec{x}}
 \propto r^{-1}e^{-F_i(r)/T}.
\]
Below the deconfining temperature, the free energy consists of a
constant-shift, L\"uscher-Coulomb, and the string-tension terms,
\(F_i(r) = \sigma_i r + c_i + \alpha_i/r\), so we should be able to
extract the string tension from its long-range part.  Indeed even with
the current limited statistics we obtain an estimate of
\(\sigma_4\)=0.102(5) for the fundamental \({\bf 4}\) representation
from our \(6\times 16^3\) lattice at \(\beta\)=10.65 (see Figure
\ref{fig:tension}).
The correlation in the antisymmetric diquark \({\bf 6}\)
representation is noisier but still yields an estimate of
\(\sigma_6\)=0.128(28).  So there seem two different string
tensions.  The higher \({\bf 10}\) and \({\bf 15}\) representations
are still too noisy to extract any tension estimates.

Conclusions: For the phase structure, we confirmed with better data
that in SU(4) pure-gauge theory the known bulk phase transition is at
\(\beta_b \sim 10.2\) and does not move with \(L_t\), and for
\(L_t=6\) the \(T\ne0\) deconfining phase change is separated from the
bulk transition and is at \(\beta_d \ge 10.75\).  A new observation is
that this \(T\ne0\) deconfining phase change does not appear stronger
than weakly first-order SU(3) deconfining transition \cite{Td}.  For
string tensions we discovered many things: We see string tension
signals for \({\bf 4}\) and \({\bf 6}\), but not yet for \({\bf 10}\)
or \({\bf 15}\).  There seem two different string tensions,
\(\sigma_4\sim 0.102(5)\) and \(\sigma_6\sim 0.128(28)\) at
\(\beta\)=10.65.  So their ratio is \(\displaystyle 1 <
\sigma_6/\sigma_4 <2\).  And combining the phase structure and tension
calculations, \(T_d/\sqrt{\sigma_4(T=0)} < T_d/\sqrt{\sigma_4(T\sim
T_d)} \sim 0.53 < \sqrt{3/\pi(d-2)}\).  We plan to accumulate more
statistics on larger and finer lattices, probably using several
motherboards of the QCDSP parallel supercomputer being built at the
RIKEN BNL Research Center.

\end{document}